# Analyzing The Mirai IoT Botnet and Its Recent Variants: Satori, Mukashi, Moobot, and Sonic

Angela Famera, Ben Hilger, Suman Bhunia, Patrick Heil

*Abstract*—**Mirai is undoubtedly one of the most significant Internet of Things (IoT) botnet attacks in history. In terms of its detrimental effects, seamless spread, and low detection rate, it surpassed its predecessors. Its developers released the source code, which triggered the development of several variants that combined the old code with newer vulnerabilities found on popular IoT devices. The prominent variants, Satori, Mukashi, Moobot, and Sonic[1], together target more than 15 unique known vulnerabilities discovered between 2014-2021. The vulnerabilities include but are not limited to improper input validation, command injections, insufficient credential protection, and out-of-bound writes. With these new attack strategies, Satori compromised more than a quarter million devices within the first twelve hours of its release and peaked at almost 700,000 infected devices. Similarly, Mukashi made more than a hundred million Zyxel NAS devices vulnerable through its new exploits. This article reviews the attack methodologies and impacts of these variants in detail. It summarizes the common vulnerabilities targeted by these variants and analyzes the infection mechanism through vulnerability analysis. This article also provides an overview of possible defense solutions.**

## I. INTRODUCTION

The ubiquitous deployment, promising services, and low-cost solutions have given the Internet of Things (IoT) immense popularity [1]–[3]. The implementation of IoT-related solutions not only has made our lives easier but has revolutionized the world economy as well. Recent economic surveys have shown that IoT devices will have an economic impact of $11 trillion per year by 2025, contributing to almost 11% of the world's overall economy. The surveys also forecast that approximately a trillion IoT devices will be deployed worldwide by 2025 [4]. With the enormous growth prospects of IoT, however, several inevitable concerns, such as service-related inconsistency and security-related issues, have drawn the attention of researchers towards IoT networks [5].

One of the security concerns includes the formation of malicious botnets by threat actors. A botnet is a network of computers connected to the Internet, called bots, that run autonomous programs. Benign botnets play a positive role by automating tasks, improving algorithms, and facilitating chat

Angela Famera, Ben Hilger, Suman Bhunia, and Patrick Heil are with the Department of Computer Science and Software Engineering, Miami University, USA, E-mail: fameraag@miamioh.edu, hilgerbj@miamioh.edu, bhunias@miamioh.edu, and heilpj@miamioh.edu

[1]Sonic is an unofficial name given by the authors as this botnet does not currently have a formal name

rooms. On the other hand, evil actors can use bots for many malicious purposes, such as financial gain, sniffing of user activities, network disruption, and the spreading of various forms of malware [6]–[9]. With smart systems intertwining with everyday household devices such as watches, cameras, and refrigerators, hackers have many IoT devices to carry out their desired tasks. Many of these IoT devices use weak credentials and suffer from vulnerabilities that allow for them to be compromised and turned into malicious bots [10], [11]. A botnet is a network of thousands or millions of bots that are commanded by a single entity.

This paper analyzes an IoT botnet known as Mirai, which emerged in 2016 and is one of the largest botnets that uses IoT devices [12]–[16]. Within its first day, Mirai infected over 65,000 IoT devices, and at its peak, it infected over 600,000 devices such as routers, air-quality monitors, and personal surveillance cameras [17]. Mirai is still used today, but since its source code was leaked, the greatest security threats have come from its emerging variants [18], [19].

In addition to analyzing Mirai, this article extensively covers its most recent variants: *Satori* (2017), *Mukashi* (2020), *Moobot* (2020), and an unnamed botnet we refer to as *Sonic* (2021), as it attacks a SonicWall Secure Socket Layer (SSL)-VPN vulnerability. The most notable property that makes these variants stand out is the vulnerabilities they attack within specific IoT targets. The four variants together target over 15 unique known vulnerabilities discovered between (2014-2021), and a handful of zero-day vulnerabilities as well. They infect a variety of devices, including but not limited to routers, switches, travel-time calculators, cameras and video surveillance products, storage devices, and video conferencing tools. These botnets hold a lot of control over IoT networks and pose a major threat to anyone, anywhere. Satori, like Mirai, for example, peaked at almost 700,000 infected devices [20]. Over 100 million Zyxel NAS devices are deployed all over the world, and Mukashi can infect these devices through the vulnerabilities it exploits [21]. During the Ukraine-Russia 2022 conflict, Mirai and Moobot were used for DDoS attacks against various government and business organizations, both in Russia and Ukraine [22]. Then Sonic (which still does not officially have a name) continues to utilize three zero-day vulnerabilities, leaving security experts ignorant.

While in-depth research on the Mirai botnet has been done in the past (more details in Section VI), to the best of our knowledge, this is the first detailed comparative study done on Mirai and how it influences current botnet variants discovered after 2019. This paper studies some of the most current IoT botnets and concludes them and Mirai to gain a better understanding of how they work and how they can be prevented. In summary, the main contributions of the current paper are as follows.

- An in-depth analysis of Mirai and four of its most recent (surfaced after 2019) variants: *Satori, Mukashi, Moobot,* and *Sonic*;
- Analyzing the source code to understand the attack behavior;
- Drawing a timeline of crucial events to understand the behavioral changes of Mirai variants;
- Comprehensive comparisons between the aforementioned variants and Mirai;
- A thorough study of common vulnerabilities and some zero-day exploits utilized by the variants; and
- Review the possible defense measures that can be taken against these variants.

The rest of this paper is organized as follows. Section II provides relevant background information on botnets and IoT devices and how these two subjects relate. Section III discusses Mirai's timeline of attacks, architecture, and propagation methods supported by fragments of the source code. Subsequently, in Section IV, we provide background on Satori, Mukashi, Moobot, and Sonic, as well as cover their attack methodologies, exploitation techniques, and similarities with Mirai in depth. Section V highlights the different defense solutions users and manufacturers should take to mitigate IoT infection. In Section VI, we summarize research papers and surveys dealing with botnets, specifically Mirai, and outline the motivations for the current paper. Finally, Section VII concludes the paper with remarks for the future scope of work.

## II. A General Overview of Botnets

This section provides valuable background information on the nature and origins of botnets and their growth in recent history. The extended forms of all abbreviations used in this paper can also be referenced in the appendix.

### A. What is a Botnet?

A botnet is a collection of bots connected to and controlled by a Command and Control (C&C) channel [23]. Bots are used all over the Internet and on various systems, such as video games and social media platforms. In what is considered a more familiar setting, bots are autonomous programs built to perform automated tasks, such as commenting on social media posts or acting as non-player characters (NPCs) in multiplayer video games. In most cases, the bots encountered daily are harmless, but they may also be programmed to attack the host device.

A malicious botnet is an attack mechanism used by cybercriminals to perform various malicious actions, including but not limited to Denial of Service (DoS) attacks, spam distribution, network scanning, exploration, and exploitation [24]. In the security world, bots are host machines, devices, and computers infected by malicious code that enslaves them to the C&C [23]. The C&C updates and guides bots to perform a desired task by acting as the communication link between bots and an individual known as a botmaster [23]. The primary purpose of the botmaster is to control the botnet by issuing commands through the C&C to perform malicious and illegal activities.

### B. Common Botnet C&C Architectures

The C&C is an important and defining characteristic of how a botnet operates. It exists in three common structures: centralized, decentralized, and hybridized. In a **centralized** architecture, there exists a central C&C giving instructions to the botnet. Whether a botmaster uses one centralized channel or many, all bots in a botnet must report to the botmaster for instructions. In a **decentralized** architecture, bots can be interconnected using Peer-to-peer (P2P) communication. This means that bots could be used to give instructions or perform specific functions to hide the C&C channels, rather than all bots reporting to a central C&C. In a **hybridized** architecture, a combination of centralized architecture, decentralized architecture, and encryption is used to hide botnet traffic. In this architecture, bots are broken up into servers and clients. Client bots block incoming connections, and server bots (which also can act as client bots) listen to predefined ports for incoming connections while using encryption to communicate with one another.

Each architecture has its advantages and drawbacks. While a centralized architecture allows the botmaster to monitor the status and distribution of the botnet, decentralized and hybridized architectures make it more difficult for security experts to detect and shut down a botnet [23].

### C. IoT devices and Botnets

IoT devices are best described as gadgets and appliances that can connect wirelessly to a network and transmit data. According to Jeremy Losaw, the author of Inventors Digest magazine, three different components make up an IoT device: 1) the ability to send/receive data; 2) the ability to establish a network connection such as Wi-Fi, Bluetooth, or a cellular connection to communicate with the cloud; and 3) the ability to manage data coming to/from the device with a back-end database or website [25].

IoT devices are popular botnet targets because if one device is found with a vulnerability, it is likely that similar devices contain the same vulnerability. Some of the most prominent vulnerabilities are Improper Input Validation, OS Command Injection, SQL Command Injection, etc. [26]. If a botnet can effectively identify those devices while evading detection, it can widen the size of its network in a matter of hours or days [27]. For example, Satori, a variant of Mirai, infected 280,000 devices within the first 12 hours of its activation.

## III. Mirai's Timeline, Architecture, and Attack Methodology

In this section, we take a deep dive into the conventional Mirai botnet and its old variants. We discuss the new variants in Section IV. First, we provide a comprehensive timeline of the events and discuss the famous Mirai attacks on OVH, Krebs on Security, and Dyn, Inc. DNS server. Next, we discuss the impact of these attacks. Finally, we provide a detailed attack methodology of the Mirai botnet and explain how the various components work together.

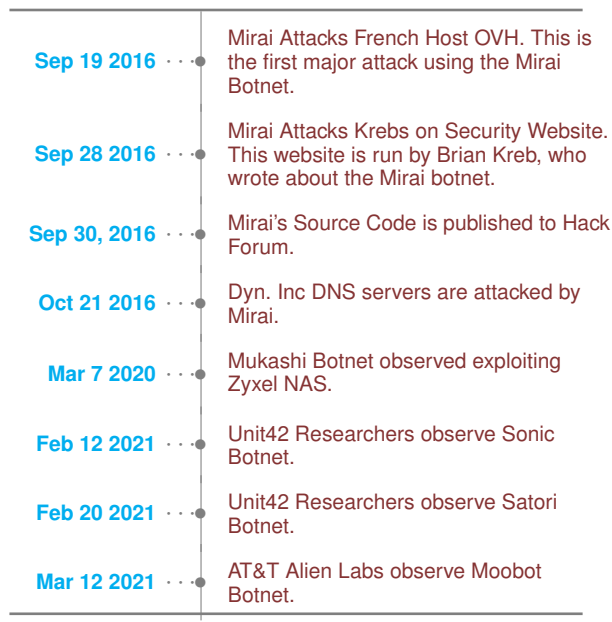

| | |
|---|---|
| **Sep 19 2016** | Mirai Attacks French Host OVH. This is the first major attack using the Mirai Botnet. |
| **Sep 28 2016** | Mirai Attacks Krebs on Security Website. This website is run by Brian Kreb, who wrote about the Mirai botnet. |
| **Sep 30, 2016** | Mirai's Source Code is published to Hack Forum. |
| **Oct 21 2016** | Dyn. Inc DNS servers are attacked by Mirai. |
| **Mar 7 2020** | Mukashi Botnet observed exploiting Zyxel NAS. |
| **Feb 12 2021** | Unit42 Researchers observe Sonic Botnet. |
| **Feb 20 2021** | Unit42 Researchers observe Satori Botnet. |
| **Mar 12 2021** | AT&T Alien Labs observe Moobot Botnet. |

Figure 1: Timeline of Mirai and botnet discovery. The first few events are discussed in Section II while the last four events are discussed in Section IV.

### A. Timeline of Events

Paras Jha and Josiah White created Mirai [28]. Jha and White co-founded Protraf Solutions, which offered mitigation services for DDoS attacks [28]. Paras Jha and Josiah White created Mirai, co-founders of Protraf Solutions, which offered mitigation services for DDoS attacks [28]. Mirai has created the basis for many botnets that exist today. This is due to the original creators releasing the source code to the world (under the name "Anna-Senpai") back in 2016 on Hackforums. Fig. 1 outlines some of the significant attacks caused by Mirai. In the subsequent section, we provide an overview of the most prominent attacks chronologically.

*1) Attack on OVH:* The first large-scale event was Mirai's attack on OVH, a popular Minecraft hosting service hosting platform. A source of inspiration for Jha's interest in hosting a Minecraft game server. Minecraft is a popular online video game where upwards of $100,000 can be earned by hosting a game server in the summer months [29]. As a result, Jha was interested in performing DDoS attacks on other Minecraft servers to attract business to his server [29], [30]. This resulted in the first major DDoS attack, which occurred on September 19th, 2016, when Mirai was used against OVH.

During the DDoS attack, Mirai used 145,000 infected devices to send 1.1 Tbps data traffic to OVH's servers, bringing their services to a halt. This amount of traffic was unprecedented in 2016, being a magnitude larger than vDOS, which maxed out at 50 Gbps [29]. vDOS was a popular DDoS-as-a-service provider and was used in the gaming industry to gain a competitive advantage against opponents. Mirai could target multiple IP addresses simultaneously, allowing it to infect an entire network rather than a specific server, application, or website [29].

*2) Attack on Krebs on Security:* Shortly after the attack on OVH, Mirai launched another DDoS attack on Brain Kreb's security website *Krebs On Security* [29]. It seemed at first that the attack was instigated as retribution for an article Krebs had posted about vDOS [29]. The article detailed a DDoS-mitigation firm that seemed to hijack web addresses believed to be controlled by vDOS [29]. In reality, Jha admitted to a friend that this attack was paid for by a customer who rented a bunch of Mirai-infected devices [31]. This DDoS attack peaked at 623 Gbps, forcing Kreb's DDoS mitigation service, Akamai, to drop Kreb's website due to the incurred costs of the attack [29]. It took four days for *Krebs On Security* to go back online. Right after this attack, Mirai's source code was released on the dark web.

*3) Attack on Dyn, Inc. DNS Server:* Soon after the attack on *Krebs On Security*, the release of the code resulted in an attack that took down Dyn. Inc's DNS servers bring down major websites on the east coast of the United States [30]. Upon the release of Mirai's source code, the hacker was able to launch an attack in October 2016 that left the US East Coast without access to major websites, such as Amazon, Netflix, PayPal, and Reddit [29]. To this day, Dyn, Inc. is still unable to assess the full weight of the assault.

### B. Impact of the Attacks

The *Krebs On Security* attack brought Mirai to the forefront of the FBI's investigation. The FBI joined with the private industry to figure out the inner-workings of Mirai. Akamai created honeypots that allowed the investigators to observe how infected devices communicated with Mirai's command-and-control servers [29]. After the *Dyn, Inc* attack, a collaboration between investigators grew, and engineers from around the world came together to discuss the threat Mirai poses. The engineers see the attack on Dyn, Inc. as a proof of concept that it can affect the entire Internet if it is not mitigated. While studying the Mirai servers, the investigators noticed a trend of the botnet targeting gaming servers, and more specifically, Minecraft servers. This led investigators to discover the original intention of Mirai, which was to target Minecraft game servers to gain a competitive edge [29].

### C. Mirai's Architecture and Propagation Methodology

The architecture of Mirai consists of three main components: a **loader**, the **compromised devices** themselves, and the **C&C** [15]. The loader provides a list of devices used to start scanning and infection activities (see Fig. 2 Step 1). Commonly known as a bootstrap loader, it is the first piece of code run in a self-starting process that is responsible for loading and executing other programs without any external input [15]. The stages of Mirai's overall infection process are visualized in Fig. 2 and discussed in the subsequent sections.

```
        (o1 == 56) ||  // 56.0.0.0/8 - US
Postal Service
        (o1 == 10) ||  // 10.0.0.0/8 -
Internal network
        (o1 == 192 && o2 == 168) ||  //
192.168.0.0/16 - Internal network
        (o1 == 172 && o2 >= 16 && o2 < 32) ||
 // 172.16.0.0/14 - Internal network
        (o1 == 100 && o2 >= 64 && o2 < 127)
|| // 100.64.0.0/10 - IANA NAT reserved
        (o1 == 169 && o2 > 254) ||  //
169.254.0.0/16 - IANA NAT reserved
        (o1 == 198 && o2 >= 18 && o2 < 20) ||
 // 198.18.0.0/15 - IANA Special use
        (o1 >= 224) ||  // 224.*.*.** -
Multicast
        (o1 == 6 || o1 == 7 || o1 == 11 || o1
 == 21 || o1 == 22 || o1 == 26 || o1 == 28
|| o1 == 29 || o1 == 30 || o1 == 33 || o1
== 55 || o1 == 214 || o1 == 215)  //
Department of Defense
    );
    return INET_ADDR(o1,o2,o3,o4);
}
```

The infected IoT devices are responsible for further spread-
ing infection on remote devices. Mirai's bots independently
scan the Internet for other devices running on Telnet by
sending out TCP SYN probes (see Fig. 2 Steps 2 & 3) targeting
ports 23 (Telnet) and 2323 (alternate port for Telnet) [15].
Telnet is an outdated network protocol administrators use that
provides a virtual connection between remote and local devices
[32].

*2) Brute Force Infection:* When a susceptible port is iden-
tified by responding to the TCP SYN probe previously dis-
cussed, the malware tries to log into the system by brute
force using ten randomly chosen username and password
combinations from a preconfigured list of around 60 frequently
used credentials by IoT devices (See Fig 2 Step 4) [7]. These
common usernames and passwords can be found in Table I.

```
// Retry
if (++(conn->tries) >= 10) {
    conn->tries = 0;
    conn->state = SC_CLOSED;
}
else {
    setup_connection(conn);
    printf("[scanner] FD%d retrying with
    different auth combo!\n", conn->fd);
}
```

*3) Reporting Back to the C&C:* All bots (e.g., compromised
IoT devices) report back to the C&C and obtain instructions
from the server (see Fig 2 Step 5). These connections to
the C&C use a plain TCP socket with an address and port
hard-coded in the source [15]. If successfully logged in, the
malware forwards the compromised device's IP and other
useful credentials to the C&C.

*4) Covering Tracks:* After reporting back to the C&C,
Mirai closes the entry point by deleting the downloaded binary
and renaming the process using a random string to keep other
malware from infecting the same device (see Fig 2 Step 6)
[33].

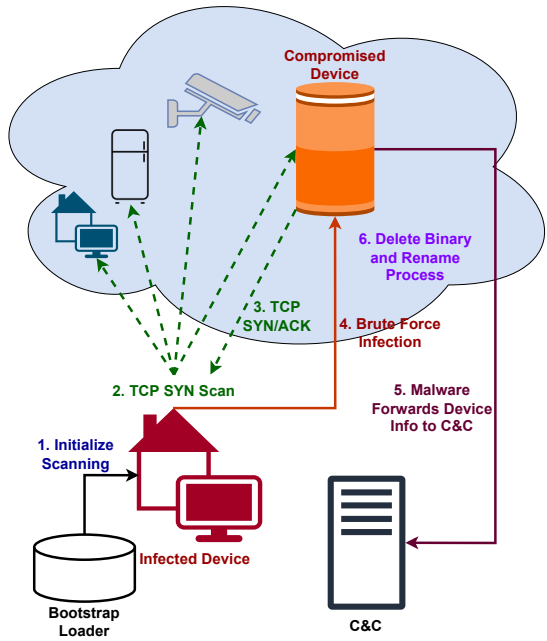

Figure 2: The four different stages Mirai and its bots use to
infect IoT devices

*1) Initial Scanning Strategy:* Mirai, along with all its vari-
ants, has a core strategy for finding and infecting IoT devices
[15]. To evade detection, Mirai uses a unique pseudorandom
number generator (PRNG) created by its developers to search
for random IPv4 addresses [15]. Another important behavior
of Mirai is that it avoids certain ports and only looks for ports
in the range [1024, 65535]. It avoids invalid addresses and
selected large organizational networks such as the Department
of Defense (DoD), the US Postal Service, General Electric
Company, and the Hewlett-Packard Company [15], [32]. It
also avoids Internet Assigned Numbers Authority (IANA)
addresses and internal networks [32]. The snippet from the
`scanner.c` file shows where these avoided ports are defined
[32].

```
// Found in scanner.c
do {
    source_port = rand_next() & 0xffff;
} while (ntohs(source_port) < 1024);

// Addresses avoided
static ipv4_t get_random_ip(void) {
    uint32_t tmp;
    uint8_t o1, o2, o3, o4;
    do {
        tmp = rand_next();
        o1 = tmp & 0xff;
        o2 = (tmp >> 8) & 0xff;
        o3 = (tmp >> 16) & 0xff;
        o4 = (tmp >> 24) & 0xff;
    }
    while (o1 == 127 ||  // 127.0.0.0/8 -
Loopback
        (o1 == 0) ||  // 0.0.0.0/8 - Invalid
address space
        (o1 == 3) ||  // 3.0.0.0/8 - General
Electric Company
        (o1 == 15 || o1 == 16) ||  //
15.0.0.0/7 - Hewlett-Packard Company
```

Table I: Usernames and Passwords used in Mirai's Brute Force Attacks

| Username | Password | Username | Password |
|----------|----------|----------|----------|
| 666666 | 666666 | root | 7ujMko0admin |
| 888888 | 888888 | root | 7ujMko0vizxv |
| admin | *(none)* | root | 888888 |
| admin | 1111 | root | admin |
| admin | 1111111 | root | anko |
| admin | 1234 | root | default |
| admin | 12345 | root | dreambox |
| admin | 123456 | root | hi3518 |
| admin | 54321 | root | ikwb |
| admin | 7ujMko0admin | root | juantech |
| admin | admin | root | jvbzd |
| admin | admin1234 | root | klv123 |
| admin | meinsm | root | klv1234 |
| admin | pass | root | pass |
| admin | password | root | password |
| admin | smcadmin | root | realtek |
| admin1 | password | root | root |
| administrator | 1234 | root | system |
| Administrator | admin | root | vizxv |
| guest | 12345 | root | vizxv |
| guest | guest | root | xc3511 |
| mother | fucker | root | xmhdipc |
| root | *(none)* | root | zlxx. |
| root | 00000000 | root | Zte521 |
| root | 1111 | service | service |
| root | 1234 | supervisor | supervisor |
| root | 12345 | support | support |
| root | 123456 | tech | tech |
| root | 54321 | ubnt | ubnt |
| root | 666666 | user | user |

## IV. Recent Variants and their Attack Methodology

Even though the original creators of Mirai were found and convicted, the public release of the source code at GitHub repository [32] allowed the botnet to grow and evolve into the many variants seen today. Variants use the same source code as the original Mirai botnet, often including additional features (such as using more recent exploits), and are launched and controlled by different actors.

The four variants we discuss in depth are Satori, Moobot, Mukashi, and Sonic. These four variants are prevalent today and take advantage of recently discovered exploits to infect IoT devices. As shown in Fig. 1, it was observed in March 2020 that Mukanashi exploited Zyxel NAS services. [34]. Then, Unit42 researchers observed the Sonic Botnet in February 2021 [35]. Later in the same year, they also observed the Satori Botnet reappearing after its initial discovery in 2017 [36]. Finally, in March 2021, AT&T Alien Labs observed the Moobot Botnet [37].

All of these variants exhibit similar characteristics to Mirai, such as scanning and propagation. The most notable difference between them all, however, is the vulnerabilities and the type of devices they attack to propagate. A summary of the variants and how they compare to Mirai can be found in Table II, and all notable vulnerabilities they use can be found in Table III. The rest of the section analyzes these four variants in detail.

### A. Satori

Satori was a Mirai variant initially active between 2017 and 2018. In December 2017, Satori infected over 280,000 IoT devices within 12 hours of its activation [27]. Satori exploited improper input validation using CVE-2014-8361 and CVE-2017-17215 to connect to Huawei routers and Realtek Software Development Kits (SDK) on ports 37215 (Huawei HG532 Service) and 52869 (Universal Plug and Play (UPnP)) [27]. See Table III for a summary of these two vulnerabilities.

Although it had supposedly dwindled since then, researchers at Palo Alto Networks found that Satori was active again. In February 2021, the researchers found the variant trying to use OS command injections (CVE-2020-9020) to exploit Iteris Vantage Velocity Field Unit 2.3.1, 2.4.2, and 3.0 devices.

*1) Iteris Vantage Velocity OS Command Injection:* The most recent Satori attack was on Iteris devices. **Iteris** is a global leader in smart mobility infrastructure management that applies cloud computing, artificial intelligence, advanced sensors, advisory services, and managed services towards safe, efficient, and sustainable mobility [54]. Vantage Velocity is a Bluetooth or Wi-Fi-based system that measures travel time. As most vehicles house a Bluetooth enabled device, such as the driver's smartphone, it senses the device's MAC address as it passes the field unit and transmits the time and location of the device to a central host system [55]. The host system then calculates the average travel times and speeds from this information. Vantage Velocity units are meant to be installed at various intersections of a road's network to capture the time a Bluetooth-enabled device passes through the intersection [55].

Attackers compromise these devices by injecting commands via HTTP requests into the `cgi-bin/timeconfig.py` file via shell meta-characters in the Network Time Protocol (NTP) Server field (CVE-2020-9020) [36], [56]. The attackers then use `wget` to download the file `arm7` from the server 198.23.238.203, then change the permissions to ensure that it could be executed by the current user [36]. This server functions as a malicious shell script that provides malware download services through HTTP port 80 and is also believed to serve as a C&C server with port 5684 [36]. This botnet was also found to have the ability to attack nine different processor platforms, which were: `ARM`, `ARM7`, `MIPS`, `PowerPC`, `sh4`, `SPARC`, `m68k`, `x86_64`, and `x86_32`. This allows the attackers to target servers with different architectures, expanding the reach of the botnet [36].

*2) Satori's Relation to Mirai:* Like Mirai, Satori scans port 23 of random hosts and attempts to log in with its embedded password dictionary when port 23 is open [36]. To infect other devices, Satori executes malicious payloads from the C&C to deploy bots on new victim devices [36]. If Satori compromises one of these devices, attackers can leak sensitive data and/or conduct DDoS attacks.

### B. Mukashi

Like Satori, what makes Mukashi stand out is the specific vulnerability it exploits. Mukashi attacks the system with a critical vulnerability in OS command injection vulnerability (CVE-2020-9054) found in ZyXel NAS devices. CVE-2020-9054 allows attackers to remotely execute malicious code into the affected system [57]. In February 2020, *Krebs On Security* alerted Zyxel that this zero-day vulnerability on their devices was being abused by attackers [58]. That same week, researchers at Palo Alto Networks detected the same vulnerability exploited and dubbed the Mirai variant "Mukashi" on March 12th, 2020 [58].

*1) ZyXel NAS OS Command Injection:* Mukashi's most prominent attack was on **ZyXel** devices. ZyXel is a Taiwanese

Table II: Comparing Various Topics Between Mirai and The Four Variants

| Variant | English Translation | Year of Discovery | Estimated Number of Bots | Number of Exploits Used | Utilizes command Injection | Utilizes Improper Input Validation | Number of Current Zero Day Vulnerabilities |
|---------|---------------------|-------------------|--------------------------|-------------------------|----------------------------|------------------------------------|---------------------------------------------|
| Mirai | "Future" | 2016 | ~493,000 | 0 | ✗ | ✗ | 0 |
| Satori | "Awakening" or "Enlightenment" | 2017 | ~280,000 | 3 | ✓ | ✓ | 0 |
| Moobot | ✗ | 2019 | ~18,705 | 6 | ✓ | ✓ | 0 |
| Mukashi | "Olden Days" or "Former" | 2020 | Unknown | 1 | ✓ | ✗ | 0 |
| Sonic | ✗ | 2021 | Unknown | 7 | ✓ | ✗ | 3 |

Table III: Vulnerability Summaries

| Name | CVE ID | NVD Base Score (CVSSv3) | Date Published | Vulnerability Type | Description | Variant |
|------|--------|-------------------------|----------------|--------------------|-------------|---------|
| Hikvision [38] | CVE-2021-36260 | 9.8 Critical | 09/22/2021 | Command Injection | Hikvision products lack proper input validation and are susceptible to command injection | Moobot |
| Realtek [39] | CVE-2014-8361 | N/A | 05/01/2015 | Improper Input Validation | Realtek SDK devices allow attackers to remotely inject and execute code in the Universal Plug and Play (UPnP) Simple Object Access Protocol (SOAP) interface using a crafted NewInternalClient request | Satori & Moobot |
| Huawei [40] | CVE-2017-17215 | 8.8 High | 03/20/2018 | Improper Input Validation | Huawei HG532 routers allow authenticated attackers to remotely send malicious packets to TCP port 37215 and execute arbitrary code | Satori & Moobot |
| Vantage [41] | CVE-2020-9020 | 9.8 Critical | 02/16/2020 | OS Command Injection | Iteris Vantage Velocity Field Unit devices (version 2.3.1, 2.4.2 and 3.0) allow attackers to remotely inject commands using crafted HTTP requests | Satori |
| NAS [42] | CVE-2020-9054 | 9.8 Critical | 03/04/2020 | OS Command Injection | ZyXEL NAS devices (version 5.21) improperly sanitize the program's username parameter, allowing unauthenticated attackers to execute arbitrary code | Mukashi |
| Tenda [43] | CVE-2020-10987 | 9.8 Critical | 07/13/2020 | OS Command Injection | The Tenda AC15 AC1900 Dual Band Wi-Fi router goform/setUsbUnload endpoint (version 15.03.05.19) allows authenticated attackers to remotely execute arbitrary commands | Moobot |
| DrayTek [44] | CVE-2020-8515 | 9.8 Critical | 02/01/2020 | OS Command Injection | DrayTek Vigor2960, Vigor3900, and Vigor300B devices allow unauthenticated attackers to remotely execute code with root privileges | Moobot |
| Grandstream [45] | CVE-2020-5722 | 9.8 Critical | 03/23/2020 | SQL Injection | The Grandstream UCM6200 HTTP (versions before 1.0.19.20) interface allows unauthenticated attackers to remotely perform SQL injections using crafted HTTP requests | Moobot |
| WIFICAM [46] | CVE-2017-8225 | 9.8 Critical | 04/25/2017 | Insufficiently Protected Credentials | Wireless IP Camera (P2P) WIFICAM devices allow attackers to bypass authentication by providing an empty loginuse and loginpas parameter in the Uniform Resource Identifier (URI) to access credentials in the .ini files | Moobot |
| VisualDoor [47] | N/A | N/A | 01/23/2021 | VPN Exploit | SonicWall Secure Socket Layer (SSL)-VPN allows unauthenticated attackers to remotely execute code as a "nobody" user via the /cgi-bin/jarrewrite.sh URL | Sonic |
| DNS-320 [48] | CVE-2020-25506 | 9.8 Critical | 02/02/2021 | OS Command Injection | The D-Link DNS-320 Revision Ax system_mgr.cgi component (firmware version is v2.06B01) allows attackers to remotely execute arbitrary code | Sonic |
| Yealink [49] | CVE-2021-27561 | 9.8 Critical | 10/15/2021 | Command Injection | Yealink Device Management (DM) (version 3.6.0.20) allows unauthenticated attackers to remotely execute arbitrary commands on the server with root privileges via the /sm/api/v1/firewall/zone/services URI | Sonic |
| Arm [50] | CVE-2021-27562 | 5.5 Medium | 05/25/2021 | Out-Of-Bounds Write | Calling secure functions under the Non-secure Processing Environment (NSPE) handler mode in Arm Trusted Firmware devices could allow attackers to trigger a stack underflow that renders the incorrect operation of the Secure code execution | Sonic |
| OBR [51] | CVE-2021-22902 | 9.8 Critical | 02/28/2021 | Code Injection | Micro Focus Operation Bridge Reporter (OBR) products (version 10.40) improperly sanitize the username parameter in the LogonResource endpoint, allowing attackers to remotely execute arbitrary code | Sonic |
| Netis [52] | CVE-2019-19356 | 7.5 High | 02/07/2020 | OS Command Injection | Netis WF2419 routers (version V1.2.31805 and V2.2.36123) allow unauthenticated attackers to remotely execute code with root privileges through the router Web management page | Sonic |
| Netgear [53] | CVE-2020-26919 | 9.8 Critical | 10/09/2020 | Other | NETGEAR JGS516PE switches (versions prior to 2.6.0.43) allow unauthenticated attackers to execute arbitrary code | Sonic |

manufacturer that has 100 million devices worldwide [58]. ZyXel NAS devices provide personal cloud storage and allow users to access their data from anywhere using a mobile device. These devices contain a pre-authentication command injection vulnerability that allows the remote execution of arbitrary code within the vulnerable device by unauthenticated attackers. This vulnerability lies within the `weblogin.cgi` program, which fails to properly sanitize the username parameter during authentication, allowing attackers to use special characters like `'` and `;` to close strings and concat commands [34]. Although the web server that runs on ZyXel devices does not run as a root user, the devices contain a `setuid` utility that can be leveraged to run commands with root privileges [59]. By sending a certain HTTP POST or GET request to `weblogin.cgi`, remote unauthenticated attackers can execute arbitrary code on the NAS device [34], [59].

*2) Mukashi's Relation to Mirai:* Similar to Mirai, Mukashi randomly scans TCP port 23 of IoT hosts and performs brute force attacks using default and previously recorded credentials to log into Zyxel NAS products, as well as digital video recorders (DVRs), security cameras, and other similar devices

[21], [57]. Once successfully connected, it sends the machine's information to the C&C server 45.84.196.75 on TCP port 34834 in the form `<host ip address>:23 <username>:<password>` [21], [60]. To evade detection, Mukashi binds to TCP port 23448 to ensure only one instance runs on the infected system [21], [61]. When malware has been executed on the infected host, Mukashi prints the message *"Protecting your device from further infection"* to the console [21], [61]. Once initialized, Mukashi notifies the C&C server 45.84.196.75 listening on TCP port 4864 that it is ready for a command from the C&C [60]. In summary, TCP port 23448 is used to run a single instance, TCP port 34834 on 45.84.196.75 is where the infected hosts' credentials are sent, and TCP port 4864 on the same server is where the commands are sent through.

Mukashi's C&C supports various commands, credentials, and attacks, all using custom encryption and decryption. These decrypted commands and attacks can be found in Table IV [61]. These are the encrypted commands used by Mukashi. The commands themselves were decrypted, and many of them were used by Mirai. The actual operation of these commands is not published although, so we assumed they were the same as

Table IV: Mukashi C&C Decoded Commands and Attacks [61]

| Commands | Description |
|----------|-------------|
| .http | HTTP flood |
| .tcp | Normal TCP scan |
| .tcpbypass | TCP scan that can bypass Firewall rules |
| .udp | UDP flood with more options |
| .udpbypass | UDP scan that can bypass firewall rules |
| .udphex | N/A |
| .udpplain | UDP flood with fewer options, optimized for higher PPS |
| .udprand | UDP flood with randomized port and packet content |
| killallbots | Code to kill all the bots under the control of C&C |
| killer | Code for killing specific service |
| PING | Ping flood |
| scanner | Code or process responsible for scanning |

Mirai's. These commands are used to perform various DDoS attacks, Mirai and other variants have used in the past.

### C. Moobot

Moobot is another recent variant of Mirai that utilizes five vulnerabilities, two of which Satori also used in its exploits in 2017 (see Table III). Moobot was first seen in 2019 by the Network Security Research Lab at 360 (360Netlab) [62], and its main targets are Docker APIs, Small Office Home Office (SOHO) devices, fiber routers, and IoT devices.

*1) Notable Attacks by Moobot:* In September 2019, 360Netlab published an article discussing various botnets monitored on 185.244.25.0/24, specifically Moobot and its variants such as `Moobot.socks5`, `Moobot.tor`, `Moobot.tor.b`, `Moobot.go`, `Moobot.tor.go`, and `Moobot.c` [63].

In July 2020, 360Netlab published an article describing the vulnerabilities of Moobot attacks, as well as its DDoS activity [62]. Moobot has been active since the start of 360Netlab's tracking attacking targets worldwide and about 100 targets per day. Moobot has attacked nearly 20,000 targets in the United States and around 48,000 IP addresses in Brazil between March 2020 and May 2020. In July 2020, Cloudflare also detected and mitigated a UDP-based DDoS attack believed to be generated by Moobot that peaked at 654Gbps [64].

In November 2020, 360Netlab, CNCERT, and Qihoo 360 published a joint article about Moobot using another zero-day vulnerability to target UNIX CCTV DVRs [65]. Later the next year, in September 2021, FortiGuard Labs reported that Moobot was using a remote code execution vulnerability (CVE-2021-36260) to infect various products from Hikvision, one of the largest video surveillance brands in the world [66].

Most recently, in February 2022, 360Netlab reported on some NTP amplification, UDP/STD/OVH floods, and other attacks on Russian and Ukrainian websites caused by Moobot, Mirai, gafgyt, ircbot, and ripprbot [22]. The attacks were launched against four 185.34.x.x/24 IPs, all belonging to Ukrainian bank `oschadbank.ua` [22].

*2) Hikvision Router Command Injection:* **Hikvision** is the world's largest manufacturer of video surveillance products and solutions and is based out of Hangzhou, China. CVE-2021-36260 is a command injection vulnerability found in many different Hikvision IP cameras and products [67]. For this vulnerability, a command is inserted into the XML payload in conjunction with the HTTP PUT request sent to the /SDK/webLanguage endpoint [68]. This results in command execution at the root user. While Hikvision did release a statement warning users about the vulnerability, devices are still susceptible if users have not updated the software.

*3) Tenda Router OS Command Injection:* Moobot was seen by AT&T scanning for vulnerabilities in **Tenda** routers. Tenda is a global supplier of networking devices and equipment based out of Shenzhen, China, whose AC15 AC1900 Dual Band Wi-Fi routers are susceptible to OS command injection (CVE-2020-10987) [69], [70]. The setUsbUnload function in the router's `bin/httpd` binary file contains a `deviceName` parameter that is passed directly to a `doSystemCmd` function [70]. This `deviceName` parameter can be set through an authenticated request, allowing attackers to execute arbitrary system commands [70]. For example, if wanting to reboot the router, an attacker can set the `deviceName=; reboot` (readers are advised to pay attention to the semicolon) to be passed to the `doSystemCmd` function [70].

*4) Draytek Switch OS Command Injection:* **Draytek** is a Taiwan-based supplier of networking devices and equipment. While currently undergoing reanalysis at the time of this writing, select beta versions of DrayTek Vigor2960, Vigor3900, and Vigor300B switches and routers are susceptible to OS command injections (CVE-2020-8515). These devices allow unauthenticated remote code execution via shell metacharacters to the `cgi-bin/mainfunction.cgi` URI [71]. This vulnerability can be used to sniff network traffic and install backdoors on the devices [72]. There are two command injection points on these devices: `keyPath` and `rtick`, both located in the /www/cgi-bin/mainfunction.cgi and its corresponding web server /usr/sbin/lighttpd [72], [73]. The keyPath is used to initiate login requests but has poor input control that make unauthorized remote command execution possible [73]. The `rtick` is used to generate a CAPTCHA image but fails to verify the incoming timestamp before generating `<rtick>.gif` [73]. When the vulnerability was discovered, the `rtick` command injection was used to create two sets of web session backdoors that wouldn't expire [73].

*5) Grandstream Networks Remote SQL Injection:* **Grandstream Networks**, headquartered in Boston, manufactures IP voice and video equipment. CVE-2020-5722 refers to the vulnerability of the HTTP interface of Grandstream UCM6200 series devices to unauthenticated, remote SQL injection attacks via crafted HTTP requests [74]. When using the UCM6200 web interface, a user can retrieve their password via the "Forgot Password" feature by entering their username [75]. The username is validated against a user's table in an SQLite database, but this query was vulnerable to a reverse shell attack in UCM6200 versions before 1.0.19.20. Until version 1.0.20.17, attackers could also introduce arbitrary HTML into the password recovery email sent to the user [75].

*6) WIFICAM Camera Insufficiently Protected Credentials:* CVE-2017-8225 refers to a vulnerability found on **WIFICAM** devices [46]. These wireless IP cameras (P2P) are Chinese web cameras that allow users to stream remotely [76]. The WIFI-CAM HTTP server is based on GoAhead, a small web server present in over 700,000 IoT devices [77]. The WIFICAM server contains `system.ini` and `system-b.ini`, which are two configuration files used to store credentials [78]. Access to these `.ini` files is not correctly checked, thus allowing an attacker to bypass authentication by providing an empty `loginuse` and `loginpas` in the URI [78]. If able to bypass authentication by providing the empty parameters in the URI, an attacker can steal credentials, FTP accounts, and SMTP accounts [78].

*7) Moobot's Relation to Mirai:* Moobot scans for DVRIP/ADB/HTTP/TELNET ports and reports its scan results back to the C&C [63]. Rather than letting the bots independently scan for new devices like with Mirai, Moobot bots perform scans together and piece the results together [37]. Like Mirai, Moobot has a hard-coded list of IPv4 addresses to avoid the same range Mirai originally did [37]. A key difference between Moobot and Mirai is that Moobot uses the hardcoded string `w5q6he3dbrsgmclkiu4to18npavj7O2f` instead of the string `abcdefghijklmnopqrstuvw012345678` used in Mirai as a seed to generate an alphanumeric string [37]. This string is a key component in generating the process name to be used during execution [37]. Moobot also uses encryption to evade string-based detection. Finally, Moobot uses `prctl()`, a Linux function that is used to control or manipulate aspects of the behavior of the calling thread or process [37]. It uses `prctl()` by hiding the process name as `/var/Sofia`, the name of a video application used on the targeted Docker API devices [37].

### D. Sonic

In February 2021, a new Mirai variant was discovered by researchers at Palo Alto Networks. At the time of this writing, it is one of the latest variants to come to light [79], and there is no official name for it. As the introduction mentions, we call this variant Sonic to keep things simple. Perhaps the most notable property of this variant is that it targets six recently discovered vulnerabilities in D-Link, Netgear, and SonicWall IoT devices, as well as three currently unidentified vulnerabilities.

*1) VisualDoor SonicWall Secure Socket Layer VPN Command Injection:* The first vulnerability the Sonic variant uses is **VisualDoor**, a SonicWall Secure Socket Layer (SSL)-VPN Remote Command Injection Vulnerability [80]. SonicWall is a security platform for cloud, hybrid, and traditional environments. The bug was discovered by Phineas Fisher, a popular hacktivist [80]. These products use an old version of Bash, which is vulnerable to ShellShock, a set of vulnerabilities used to gain higher privileges and unauthorized access within Bash. SonicWall devices are therefore vulnerable to unauthenticated remote code execution (as a "nobody" user) via the `/cgi-bin/jarrewrite.sh` URL [80].

*2) D-Link DNS-320 OS Command Injection:* The D-Link **DNS-320** Revision Ax is susceptible to OS command injections in the `system_mgr.cgi` component (CVE-2020-25506) [81]. D-Link is a company based in Taiwan that manufactures network equipment, and DNS-320 is a storage device produced by D-Link that features a built-in Web File and File Transfer Protocol (FTP) server. Arbitrary command execution is caused more specifically by improper sanitation of HTTP parameters in the `f_ntp_server` [60].

*3) Yealink Device Management Command Injection:* **Yealink** is a global brand that specializes in videoconferencing tools [82]. CVE-2021-27561 is a vulnerability where Yealink Device Management 3.6.0.20 allows command injections as root via the /sm/api/v1/firewall/zone/services URI without requiring any authentication [83].

*4) ARM Out of Bounds Write:* CVE-2021-27562 is a vulnerability where in **ARM** Trusted Firmware M through 1.2, the Non-Secure world may trigger a system halt, an overwrite of secure data, or the printing out of secure data when calling secure functions under the Non-Secure Processing Environment (NSPE) handler mode [84]. Arm Trusted Firmware is an organization that provides open-source secure software that complies with ARM specifications.

*5) Micro Focus OBR Code Injection:* CVE-2021-22502 is a remote code execution vulnerability in Micro Focus **Operation Bridge Reporter (OBR)** products, affecting version 10.40 [85]. This vulnerability could allow remote code execution on the OBR server [85]. Micro Focus is a worldwide enterprise software provider that delivers mission-critical technology and supporting services to manage the IT elements of a business. Operations Bridge is a Micro Focus product used for cloud monitoring.

*6) Netis OS Command Injection:* CVE-2019-19356 describes that **Netis** WF2419 is vulnerable to authenticated Remote Code Execution (RCE) as root through the router web management page in firmware version V1.2.31805 and V2.2.36123 [86]. It is possible to execute system commands as root through the tracert diagnostic tool because of the lack of user input sanitization [86]. Tracert is used to trace the path an IP packet takes to its destination. Netis develops networking products, and WF2419 is a wireless 802.11n router.

*7) Netgear Lack of Access Control:* CVE-2020-26919 is a vulnerability where **NETGEAR JGS516PE** devices before version 2.6.0.43 are affected by lack of access control at the function level [87]. Netgear is a multinational computer networking company, and JGS516PE is a Smart Managed Plus Power over Ethernet (PoE) Switch designed for desktop or rackmount.

Sonic also uses three unknown vulnerabilities, all of which are some type of command injection.

*8) Sonic's Relation to Mirai:* Upon successful exploitation, `wget` is used to download a malicious shell script that executes several Mirai binaries one by one [60], [79].

One of the binaries is `lolol.sh`, which deletes key folders in the target machine responsible for scheduling jobs and starting up processes. It also downloads dark binaries based on the Mirai codebase that is used for propagation via the exploits mentioned above or brute forced `ssh` connections using hard-coded credentials [60]. These dark binaries are saved to a file called `nginx` for evasion since NGINX is a widely known open source web service software [60]. `lolol.sh` is also responsible for implementing several packet filters to block incoming traffic to commonly used SSH, HTTP, telnet, and other similar ports to try and make it more difficult for administrators to maintain the system [60]. lolol.sh is also supposed to be rerun every hour, but the cron command-line utility is said to be improperly configured [60].

`install.sh` is another script installed by the malware and is used to download `GoLang v1.9.4`, as well as the GoLang standard SSH and Zmap packages onto the device [60]. Go is an open-source programming language created and supported by Google, and Golang is another common way of referring to the language. The Golang SSH package implements an SSH client and server, and Zmap is a single-packet network scanner.

Two other binaries downloaded are `nbrute` and `combo.txt` [60]. `combo.txt` is a plain text file containing commonly used credentials, while `nbrute` uses `combo.txt` to brute force SSH connections with IP addresses [60]. In summary, this variant downloads binaries to schedule jobs, makes filter rules, carries out brute force attacks, and spreads malware [60].

## V. Defense Against Mirai and its Variants

Now that we have discussed attack methodologies for the old variants of Mirai let us take a look at the various defense solutions that can be used to protect IoT devices against Mirai and its variants.

### A. Change Default Passwords

One of the easiest and probably most practical ways to defend against these botnets is by changing the default username and password on IoT devices. Mirai's dictionary attack successfully penetrated hundreds of thousands of devices and only tested a handful of weak credentials.

IoT devices are produced in abundance by different manufacturers all over the world. When produced, IoT devices may come with one of the following [88]:

- A factory default username and password printed on the device or in the instruction manual
- A randomized password is given to each user if the default password is not changed
- The same username and password that comes with the other devices produced by the manufacturer (take the combo "admin" and "admin" for example)

Whether a user logs into the device manually, through a web browser, or through an app, it is important that users make time to change their passwords to something more complex and unique. If IoT devices do not come with a password (e.g., Amazon Alexa, Google Hub, etc.), the device likely connects to a home router or your phone [88]. Users can protect themselves this way by making a secure PIN for their phone or securing their router login details [88]. IoT manufacturers should also take steps to randomize the usernames and passwords for their products.

### B. Vulnerability Research and Management

Antonakakis et al. predicted in 2017 that *"attacks of the future will evolve to target software vulnerabilities in IoT devices, much like the early Code Red and Confickr worms"* [7]. We saw in Section IV that what makes many of these variants unique is their ability to take advantage of a specific set of vulnerabilities in IoT devices. For example, Satori uses an OS command injection to infect Vantage Velocity Devices, while Mukashi uses an OS command injection to infect ZyXEL NAS devices. When purchasing devices, users should investigate the product, the manufacturer, and any associated vulnerabilities or security incidents. Manufacturers should take the time to program proper constraint validation, as many of these vulnerabilities are caused by injection attacks. As new vulnerabilities emerge, IoT devices should be programmed to update their firmware and operating systems automatically. Automatic updates require a modular software architecture in the event of an update failure and a PKI infrastructure to support trusted updates [7]. Hackers can not only patch vulnerabilities as they arise, but botnets also have difficulty maintaining control of an infected device once it has been rebooted [15].

### C. Other Defense Solutions

Other ways to defend devices against IoT botnets include:

- Periodically updating the firmware and the applications running on the IoT devices. Manufacturers release patches right after a vulnerability is discovered. Installing security updates/patches prevents further infection spread.
- Monitoring IP ports 2323/TCP and 23/TCP for attempts to gain unauthorized control using Telnet [89];
- Disabling Universal Plug-and-Play (UPnP) [89], [90];
- Insist on separate registration and authentication for each IoT device. Manufacturers could produce IoT devices that require consumers to register email addresses to facilitate the communication of alerts and issues [89];
- Securing the network by using VPNs, creating a separate network, and using a third-party firewall or Intrusion Detection/Prevention System [90];
- Leaving the devices offline if they are not being used for long periods [89].

## VI. Related Work

This section presents a detailed study of surveys already published in this domain and motivates the current paper. Numerous studies have been conducted on Mirai and botnets in general in the past decade. Table V provides an overview of articles reviewing botnets, Mirai, and its variants. The following is the description of each column in Table V:

- **Year of Publication:** The year the paper was published
- **Title:** Title of the paper
- **Analyze Mirai's Infrastructure:** Describes Mirai's architecture (i.e. how the botnet works)
- **Included Mirai Source Code:** Presents and/or discusses Mirai's source code
- **Analyzes Mirai's Attack Methodology:** Discusses Mirai's infection process
- **Mirai's Impact and Variant Relations:** Discusses the consequences or effect Mirai has had on the economy, the security industry, and/or other variants
- **Investigates Mirai Variants:** Names and describes variants of Mirai
- **Name of Mirai Variants:** Lists the names of the Mirai variants
- **Defense Strategies Discussed:** Describes defense mechanisms against Mirai or botnets in general

Table V: Summary of the Related papers covering different aspects of Botnet and Mirai

| Year of Publication | Title | Analyze Mirai's Architecture | Includes Mirai Source Code | Analyzes Mirai's Attack methodology | Mirai's Impact and Variant Relations | Investigates Mirai Variant | Names of Mirai Variants | Defense Strategies Discussed | Summary |
|---|---|---|---|---|---|---|---|---|---|
| 2021 | Survey on Botnet Detection Techniques: Classification, Methods, and Evaluation [14] | ✗ | ✗ | ✗ | ✗ | | ✗ | ✓ | Survey on important/recent botnet detection efforts. Proposes a bot detection evaluation system known as CBDES. It discusses Honeypot Analysis, Communication Signature Detection, Anomaly Detection, Deep Learning, Complex Networks, Swarm Intelligence, Statistical Analysis, Distributed Approaches, and Multidimensional Detection Methods. |
| 2021 | The Circle of Life: A Large-Scale Study of The IoT Malware Lifecycle [91] | ✗ | barely | ✗ | ✓ | ✓ | barely | ✓ | Detailed analysis of the lifecycle of IoT malware and comparison with traditional malware. Presents a large-scale measurement of Linux-based IoT malware samples. |
| 2020 | Examining Mirai's Battle over the Internet of Things [15] | ✓ | ✓ | ✓ | ✓ | ✓ | General overview of 39 variants | ✗ | Comprehensive overview of the battle for and reinfection process of IoT devices by Mirai and its variants. Provide the "first epidemiological quantification of Mirai." |
| 2020 | New Variants of Mirai and Analysis [18] | ✓ | ✓ | ✓ | ✓ | ✓ | No; Authors Propose New Mirai Variants | ✗ | Studies Mirai and proposes new Mirai Variants |
| 2020 | Systematic Literature Review on IoT-Based Botnet Attack [16] | ✗ | ✗ | ✗ | ✗ | ✗ | ✗ | ✗ | Systematic literature review on IoT-based botnet attacks and detailed analysis and discussion of its primary studies |
| 2020 | Analyzing Variation Among IoT Botnets Using Medium Interaction Honeypots [10] | ✓ | ✓ | ✓ | ✓ | | General overview of variants | ✗ | Provides discussion of Mirai's functionality and focuses on how much Mirai has been modified. Provides an idea of the amount of variation present in Mirai attacks using Cowrie SSH/Telnet honeypot |
| 2020 | IoT Botnet Forensics: A Comprehensive Digital Forensic Case Study on Mirai Botnet Servers [11] | ✓ | ✗ | ✓ | ✓ | Barely | | ✓ | A comprehensive digital forensic case study on Mirai. Sets up a fully functioning Mirai botnet network architecture and conducts a comprehensive forensic analysis on the Mirai botnet server. Discussed physical and remote forensic techniques to examine the Mirai botnet server. |
| 2019 | 10 years of IoT Malware: a Feature-Based Taxonomy [92] | ✗ | ✗ | ✓ | ✓ | ✗ | ✗ | ✓ | Identification of characteristic features of several malware. Comparison of 16 of the most widespread IoT malware programs based on these features. Novel graphic representation of the malware relationships. |
| 2019 | Issues and challenges in DNS based botnet detection: A survey [13] | ✗ | ✗ | ✗ | ✗ | ✗ | ✗ | ✓ | Analysis of various Botnet detection techniques employing DNS Protocol. Thoroughly analyzes more than 200 papers and categorized the detection techniques, and proposes attributes of a Smart DNS-based botnet detection system. This paper discussed DNS-Based Detection. |
| 2018 | DDoS-Capable IoT Malwares: Comparative Analysis and Mirai Investigation [6] | ✓ | ✓ | ✓ | ✓ | No; discusses variants of DDoS-capable malware though | ✗ | ✗ | Provides taxonomy of DDoS attacks and different types of network architectures used to carry them out. Discusses DDoS-capable IoT malware. Provides Mirai background, modes of operation, source code analysis |
| 2018 | Tracking Mirai Variants [19] | ✗ | ✗ | ✗ | ✓ | ✓ | General overview of multiple variants with an emphasis on Masuta, Owari, and Wicked | ✗ | Analyzed over 32,000 Mirai samples and developed a set of variant classification and tracking schemes. Emphasized how the data were extracted and used to classify and track Mirai variants |
| 2018 | Understanding linux malware [93] | | | | | ✓ | | | The paper fills the gap between the malware industry and academia. Security experts write blog posts to publish their findings, while the academic community requires systematic studies. |
| 2018 | Iot malware: Comprehensive survey, analysis framework and case studies [94] | ✓ | ✓ | ✓ | ✓ | ✓ | ✗ | ✓ | A generic overview of IoT Malware and the major variants. |
| 2017 | Understanding the Mirai Botnet [7] | ✓ | ✗ | ✓ | ✓ | ✓ | General overview of 33 variants | ✓ | Tracking the botnet's composition, evolution, and DDoS activities from August 1, 2016, to February 28, 2017, the researchers provide what they call the "first comprehensive analysis" of Mirai. A thorough description of Mirai, as well the botnets' timeline of events, structure and propagation, malware phylogeny and its relation to BASHLITE, types of devices infected, types of attacks and targets, and possible defense strategies IoT companies can take against botnets. The paper also discussed Security Hardening, Automatic Updates, Notifications, Facilitating Device Identification, Defragmentation, and End-Of-Life. |
| 2017 | A survey of botnet detection based on DNS [8] | ✗ | ✗ | ✗ | ✗ | ✗ | ✗ | ✓ | Describes the botnet life cycle and gives a comprehensive overview of botnet detection based on DNS traffic characteristic. A concise focus on DNS-based detection solutions or techniques. The paper also discussed Honeynet-Based, IDS/Anomaly Based, Statistical-Based, Graph-Based, Clustering-Based, Entropy-Based, Decision Tree-Based, and Neural Network-Based DNS Detection. |
| 2017 | The Mirai Botnet and the IoT Zombie Armies [9] | ✓ | ✗ | ✓ | ✓ | ✓ | General Overview; BirckerBot, Hajime | ✓ | Provides a comprehensive but succinct analysis on the Mirai botnet, its variants, and the repelling tactics and countermeasures. The paper discussed Block TCP ports used for probing and brute-forcing the device. Drop TCP egress connections containing attack traffic. Closing and stopping nonessential ports and services running on the device. Isolate an organization's intranet. Allowing Busybox execution only by a specific user. Disabling UPnP. Update devices with patches and bug fixes. |

- **Summary:** Summary of the work

A few important papers are discussed below.

Antonakakis et. al in [7] provide a comprehensive overview of Mirai. Tracking the growth, composition, evolution, and DDoS activities from August 1, 2016, to February 28, 2017, the researchers provide what they call the *"first comprehensive analysis"* of Mirai. The researchers give a thorough description of Mirai, as well as the botnets' timeline of events, structure, and propagation, malware phylogeny and its relation to BASHLITE, the types of devices infected, types of attacks and targets, and possible defense strategies IoT companies can take against botnets.

Liu and Wang in [19] analyzed more than 32,000 Mirai samples and developed a set of variant classification and tracking schemes. A heavy emphasis is placed on how the researchers extracted the data and used it to classify and track Mirai variants. The researchers take a more holistic approach to discuss the variants and provide details specifically on Masuta, Owari, and Wicked.

Griffioen and Doer provide a comprehensive overview in [15] of the battle for and reinfection process of IoT devices by Mirai and its variants. The researchers analyzed how 39 variants infect and retain control over IoT devices and provide a more detailed comparative analysis between Mirai, Miori,

Akuma, Masuta, Josho, MM, and Objorn. Mirai's system infrastructure, source code, PRNG, and interaction with its variants are described in depth.

Mirai, along with botnets in general, has been widely studied in the literature. Many of the papers shown in Table V discuss Mirai's architecture, code, and attack methodology in depth. Many of these papers also discuss defense strategies against Mirai and other botnets. However, what these papers lack is an in-depth analysis of variant attack mechanisms. While multiple papers talk about variants of Mirai as a whole, they fail to cover specific variants and analyze them in depth. To the best of our knowledge, our paper is the most recent in-depth analysis of the attack strategies of four variants discovered between 2017-2021 (i.e., Satori, Mukashi, Moobot, and Sonic) and how they compare to Mirai.

The authors in [91] methodically analyze the life cycle of IoT malware and compare it with traditional malware to examine the efficacy of current defenses against IoT malware. With an extensive measurement comprising over 166K Linux-based IoT malware samples accumulated over a year spanning six different system architectures. The authors compare their results with previous studies on traditional malware and conclude that adequate defense technology is available that needs to be put into practice.

## VII. Conclusion

In this paper, we examine Mirai and four of its recent variants. We saw that these variants are similar to Mirai in that they target similar devices, pseudorandomly search for open ports, and use brute-force attacks with a list of commonly used usernames and passwords to gain access to a device. One of the biggest differences between Mirai and these variants is that they target new vulnerabilities that the original botnet did not take into consideration. As botnets exploit these arising vulnerabilities and take over IoT devices, the compromised devices can be used for various detrimental attacks such as DDoS attacks. This costs companies copious sums of money as services are made unavailable to the public, and personal data becomes compromised. One of the best defenses against botnets is to change the default username and password so that the authentication is harder to crack. Keeping up with publicly known vulnerabilities and ensuring all files and features of a program, device, or service are immune to injections used for remote privilege escalation are other defenses. IoT devices are only becoming more popular, so greater emphasis must be placed on the security of these devices to ensure that they are not vulnerable to Mirai and its variants. Our comprehensive study of these four variants will potentially provide security experts and developers of IoT devices with new information on how to counter botnet attacks.

## Appendix

Table VI provides a list of all the abbreviations used in the paper.

Table VI: Common Abbreviations Used in Paper

| Abbreviation | Full Form |
| --- | --- |
| ADB | Apple Desktop Bus |
| ARC | Argonaut RISC Core |
| CVE | Common Vulnerabilities and Exposures |
| C&C | Command and Control |
| DDoS | Distributed DoS |
| DM | Yealink Device Management |
| DoD | Department of Defense |
| DoS | Denial of Service |
| DVR | Digital Video Recorder |
| FBI | Federal Bureau of Investigation |
| FTP | File Transfer Protocol |
| Gbps | Gigabytes per Second |
| HTML | Hypertext Markup Language |
| HTTP | Hypertext Transfer Protocol |
| IANA | Internet Assigned Numbers Authority |
| IDS | Intrusion Detection System |
| IoT | Internet of Things |
| IP | Internet Protocol |
| IPS | Intrusion Prevention System |
| IPv4 | Internet Protocol version 4 |
| IRC | Internet Relay Chats |
| MAC | Media Access Control |
| NAS | Network Attached Storage |
| NPC | Non-player Characters |
| NSPE | Non-secure Processing Environment |
| NTP | Network Time Protocol |
| NVD | National Vulnerability Database |
| OBR | Micro Focus Operation Bridge Reporter |
| OS | Operating System |
| PoE | Power over Ethernet |
| PRNG | Pseudo-random Number Generator |
| P2P | Peer-2-Peer |
| RCE | Remote Code Execution |
| RNG | Random Number Generator |
| RSA | Rivest, Shamir, Adleman |
| SDK | Software Development Kit |
| SSH | Secure Shell |
| SSL | Secure Sockets Layer |
| SOAP | Simple Object Access Protocol |
| SOHO | Small Office Home Office |
| TCP | Transmission Control Protocol |
| UPnP | Universal Plug and Play |
| URI | Uniform Resource Identifier |
| VPN | Virtual Private Network |
| 2FA | Two-Factor Authentication |

## Declarations

- **Funding** The current research is not funded by any grant.
- **Conflict of interest/Competing interests** The authors have no conflicting financial or non-financial interests.